\documentclass[twocolumn,amsmath,amssymb,aps,pra,superscriptaddress,longbibliography]{revtex4-2}
\usepackage{dcolumn}
\usepackage{bm}
\usepackage{mathtools, cases}
\usepackage[unicode]{hyperref}
\usepackage{graphicx}
\hypersetup{colorlinks=true,linkcolor=blue,citecolor=blue,filecolor=black,urlcolor=blue}

\begin{document}
\title{
Anderson localization of excitations in disordered Gross--Pitaevskii lattices
}

\author{Yagmur Kati}
\email{ygmrkati@gmail.com}
\affiliation{Center for Theoretical Physics of Complex Systems, Institute for Basic Science, Daejeon 34051, Korea}
\affiliation{Basic Science Program, Korea University of Science and Technology (UST), Daejeon 34113, Republic of Korea}
\author{Mikhail V. Fistul}
\email{Mikhail.Fistoul@ruhr-uni-bochum.de}
\affiliation{Center for Theoretical Physics of Complex Systems, Institute for Basic Science, Daejeon 34051, Korea}
\affiliation{Theoretische Physik III, Ruhr-University Bochum, Bochum 44801 Germany}
\affiliation{National University of Science and Technology "MISIS", Moscow 119049 Russia}
\author{Alexander Yu. Cherny}
\email{alexander.yu.cherny@gmail.com}
\affiliation{Center for Theoretical Physics of Complex Systems, Institute for Basic Science, Daejeon 34051, Korea}
\affiliation{Bogoliubov Laboratory of Theoretical Physics, Joint Institute for Nuclear Research, 141980, Dubna, Moscow region, Russia}
\author{Sergej Flach}
\email{sflach@ibs.re.kr}
\affiliation{Center for Theoretical Physics of Complex Systems, Institute for Basic Science, Daejeon 34051, Korea}
\affiliation{Basic Science Program, Korea University of Science and Technology (UST), Daejeon 34113, Republic of Korea}


\date{\today}
\begin{abstract}
We examine the one-dimensional Gross--Pitaevskii lattice at zero temperature in the presence of uncorrelated disorder. We obtain analytical expressions for the thermodynamic properties of the ground state field and compare them with numerical simulations both in the weak and strong interaction regimes. We analyze weak excitations above the ground state and compute the localization properties of Bogoliubov-de Gennes modes. In the long-wavelength limit, these modes delocalize in accordance with the extended nature of the ground state. For strong interactions, we observe and derive a divergence of their localization length at finite energy due to an effective correlated disorder induced by the weak ground state field fluctuations. We derive effective strong interaction field equations for the excitations and generalize to higher dimensions.
\end{abstract}

\maketitle

\section{Introduction}
Trapped ultra-cold atoms
are used
 for many years as an ideal playground to study
the properties of a variety of model classes of interacting
bosons in the presence of external
potentials  \cite{Pitaevskii16:book,pethick_smith_2008,RevModPhys.71.463,leggett99,fetter01:vortices}. Periodic optical potentials, obtained with the counter-propagating laser beams \cite{PhysRevLett.81.3108}, allow for emulating the physics of interacting particles in almost arbitrarily tunable crystal fields. Artificially created random potentials enable the direct observation of Anderson localization \cite{billy08,roati08,deissler10}.
Such systems allow for the observation of Bose-Einstein condensation
of ultra-cold atoms. The dynamics of the Bose-Einstein condensate is
successfully described by the Gross--Pitaevskii (GP)
model \cite{Gross_61,Pitaevskii_61}.

The thermodynamic properties of the GP lattice model \cite{Rasmussen2000,Polkovnikov2002} are determined by two conserved quantities \cite{Rasmussen2000,mithun2018weakly}: the particle density $a$ and the energy density $h$. The two relevant energy scales are  the interaction energy
$ga$ (here $g$ is the interaction induced nonlinearity strength $g$) and the kinetic energy $J$ due to the tunneling between adjacent lattice sites.  In Ref.\cite{Rasmussen2000} the complete phase diagram $h$-$a$ has been obtained for a one-dimensional GP chain in the absence of disorder. Various peculiar effects, e.g. a dynamical glass behavior \cite{mithun2018weakly},  non-Gibbsian phases \cite{Rasmussen2000}, and long-lived spatially localized rare fluctuations forming at high temperatures \cite{iubini2019dynamical},  have been predicted and observed in  extensive simulations.

An ordered GP model has a spatially homogeneous ground state. Weak excitations above that ground state result in
spatially extended Bogoliubov-de Gennes modes \cite{bogoliubov1947theory}. Their properties have been studied for spatially continuous systems using the Bogoliubov-de Gennes (BdG)  theory \cite{bogoliubov1947theory,Bogolyubov_59,de2018superconductivity,Fetter1971}.  The spectrum of these excitations was verified in experiments with atomic Bose-Einstein condensates (see the reviews \cite{RevModPhys.71.463,leggett99}).
The interplay of nonlinearity, discreteness of the media, interaction, and spatial disorder result in novel phenomena.
Disorder introduces an additional energy scale $W$. It induces long-time relaxation dynamics and glass phases, the Anderson localization \cite{Anderson58},
and the Lifshits "glass" phase  \cite{LuganPRL98,Scalettar,LuganPRL2007,Kati2020,Deng}, to name a few. The presence of interactions strongly influences Anderson localization (see, e.g., the reviews \cite{sanchez10:rev,modugno10:rev,Shapiro_12}). For bosons, repulsive interactions tend to delocalize excitations
 \cite{sanchez10:rev}.
A number of previous studies have analyzed the long-wavelength properties of BdG excitations in one-dimensional disordered GP models
\cite{Chalker,Pavloff,Fontanesi_PRA,Fontanesi_PRL,Huber,Kramer2003} in order to address superfluidity and various phase transitions.

In this work, we address the properties of the one-dimensional disordered GP lattice by going beyond the long-wavelength limit.
We compute the ground state dependence $h(a)$ both in the weak interaction ($ga \ll J,W$) and the strong interaction ($ga \gg J,W$) regimes.
The weak interaction ground state is characterized by rare regions with nonzero particle density separated by large empty parts due to
the Lifshits tail states. The strong interaction ground state is almost homogeneous with weak fluctuations induced by the disorder field.
We compute the ground state using efficient numerical schemes and find quantitative agreement with analytical approximations.
We then proceed with computing the localization properties (localization length $\xi$ and participation number $P$) of BdG excitations.
Exact numerical diagonalization confirms the divergence of $\xi$ and $P$ in the long-wavelength limit. This is due to the fact that
the BdG zero-energy eigenmode becomes a copy of the ground state in that limit.
We then report on a surprising anomalous enhancement of $\xi$ and $P$ of BdG modes at finite energy
in the strong interaction regime.
We aim at an analytical description by performing a systematic perturbation approach using first the exact BdG equations and an approximate
ground state field. We then proceed with approximating the equations as well and obtain the disordered BdG equations in the leading order for the strong interaction regime. These equations reduce the eigenvalue problem to a one-dimensional chain with short-range correlated disorder, for which the localization length can be computed analytically. We find full agreement of our results between all stages of approximations.
Most importantly, the analytical solution yields a full divergence of the localization length at finite energy.
We conclude with a
discussion of possible generalizations to higher dimensional lattice cases.

The paper is organized as follows. In Sec. II we introduce the model.   In Sec. III we study the ground state  properties by combining the analytical and numerical approaches. In Section IV we compute the localization properties of the BdG excitations. We conclude with Section V.

\begin{figure}[tp]
\centering
\includegraphics[width=0.5\textwidth]{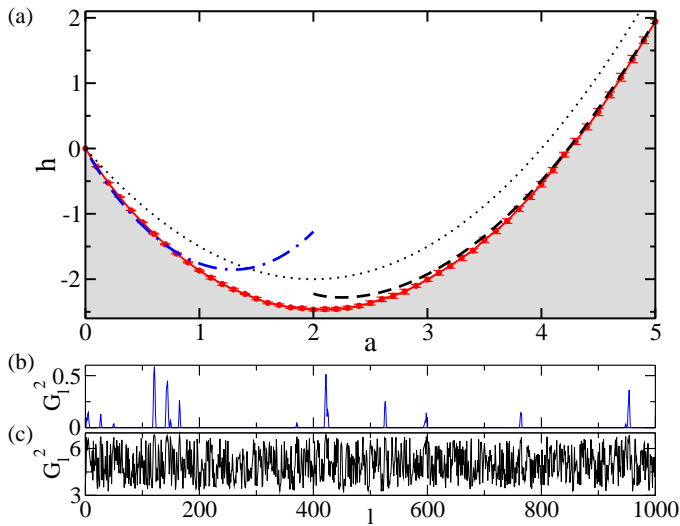}
\caption{(a) The ground state energy density $h$ versus norm density $a$.
Red circles connected with lines are the results of the numerical calculation for $W=4$, $N=10000$, $N_r=50$, and the error bars are the corresponding standard deviations due to disorder realization averaging.
The dotted black line corresponds to the ordered case $W=0$.
Dashed-dotted blue line: weak interaction approximation Eq.~(\ref{eq:hfistul}).
Dashed black line: strong interaction approximation Eq.~(\ref{eq:h_large_a}).
(b) $G_{\ell}^2$ versus $l$ for one disorder realization with $a=0.01$ and $W=4$.
(c) $G_{\ell}^2$ versus $l$ for one disorder realization with $a=5$ and $W=4$.
Note that $g=1$, $J=1$ for all cases.
}
\label{fig:phasediagram}
\end{figure}

\section{The Model}

We consider a disordered Gross--Pitaevskii (GP) model on a one-dimensional lattice. The model describes e.g. the properties of a Bose-Einstein condensate of ultracold atoms loaded onto an optical lattice formed by counter-propagating laser beams. The dynamics of such a system is governed by the Hamiltonian
\begin{equation}\label{eq:H}
\mathcal{H}=\sum_{\ell=1}^N \frac{g}{2}|\psi_{\ell}|^4 + \epsilon_{\ell} |\psi_{\ell}|^2 -J (\psi_{\ell} \psi_{\ell+1}^* +\psi_{\ell}^*\psi_{\ell+1}).
\end{equation}
$\psi_\ell$ is the condensate wavefunction amplitude on lattice site $1 \leq \ell \leq N$.
The nonlinearity parameter $g$ is related to the strength of two-body interactions in the condensate on the same site, and $J$ is the hopping strength between neighboring sites, which is related to the tunneling rate of a single particle. The on-site energies $\epsilon_\ell$
describe the spatial disorder imprinted into the system, and take random uncorrelated values
with a probability distribution function $\rho(\epsilon)$ being constant in the region $[-W/2,W/2]$
and zero outside. Hence, its mean value and variance are given by $\langle\epsilon_\ell\rangle =0$ and $\langle\epsilon_\ell^2\rangle =W^2/12$, respectively.

The Hamiltonian equations of motion
$\dot{\psi_\ell}=\partial \mathcal{H}/\partial(i{\psi_\ell}^*)$
result in
\begin{equation}\label{eq:eom}
i\dot{\psi_{\ell}}=\epsilon_{\ell} \psi_{\ell} +  g|\psi_{\ell}|^2\psi_{\ell}-J(\psi_{\ell+1}+\psi_{\ell-1}).
\end{equation}
The discrete GP equation possesses two integrals of motion:
the total energy $\mathcal{H}$ and norm $\mathcal{A}=\sum_\ell |\psi_\ell|^2$.
The energy can be measured in units of $J$, and  the uniform rescaling of the average norm $a=\mathcal{A}/N$ tunes the nonlinearity $g$. Therefore we fix $J=1$ and $g=1$ for
the numerical computations to come, but keep them explicitly in all analytical expressions.

\section{Ground state properties}

The dynamics of the system depends on the two energy and norm densities $h=\mathcal{H}/N$ and $a = \mathcal{A}/N$. For a fixed value of $a$, the GP model has a ground state (GS) of minimum energy which is characterized by the lowest possible value of $h$.
The ground state can be obtained using the method of Lagrange multipliers:
\begin{equation}\label{eq:lagrange}
\mathcal{L}=\mathcal{H}-\mu \mathcal{A}\; , \qquad \frac{\partial \mathcal{L}}{\partial{\psi_\ell}^*}=\frac{\partial \mathcal{L}}{\partial{\psi_\ell}}=0.
\end{equation}
It follows from Eqs. (\ref{eq:H}) and (\ref{eq:lagrange})
\begin{equation}\label{eq:eq4}
    \mu \psi_{\ell} =
\epsilon_{\ell} \psi_{\ell} +  g|\psi_{\ell}|^2\psi_{\ell}-J(\psi_{\ell+1}+\psi_{\ell-1}),
\end{equation}
where $\mu$ is the chemical potential (Lagrangian multiplier).
Comparing Eqs. (\ref{eq:eom}) and (\ref{eq:eq4}) yields
\begin{equation}\label{eq:psilt}
    \psi_{\ell}(t)=G_\ell e^{-i\mu t},
\end{equation}
where  $G_\ell=\psi_{\ell}(0) \geq 0$ can be chosen to be
real and non-negative.
We then obtain the set of equations that determines the ground state amplitude field $G_\ell$
\begin{equation}\label{eq:a_generalrule}
G_\ell^2=(\mu -\epsilon_\ell+J\zeta_\ell)/g \;,   \;\;\;\; \zeta_\ell=(G_{\ell+1}+G_{\ell-1})/G_\ell
\end{equation}
with the ground state correlation field $\zeta_{\ell}$.

We note that the ground state solution for the ordered case $W=0$ is simply $G_{\ell}^2=a$. Using Eq.~(\ref{eq:H})
we arrive at the analytical dependence $h=ga^2/2-2a$, which is shown as a dotted black line in Fig.~\ref{fig:phasediagram}a.

In the presence of disorder, there are three competing energy density scales:
the kinetic $J$, disorder $W$, and interaction $ga$ energy densities, respectively.
For all studied cases we use $W=4$. We distinguish the regime of weak interaction $ga \ll W$ and strong interaction $ga \gg W$.
By fixing the norm density, we numerically minimize the energy by varying the real and nonnegative variables $G_\ell$ for a given disorder realization as in Ref.\cite{Kati2020} (see also Appendix \ref{sec:numGS}). The resulting ground-state density distribution $G_{\ell}^2$ is plotted in Fig.~\ref{fig:phasediagram} for two different norm densities:  $a=0.01$ and $a=5$, with one and the same disorder realization. For each outcome, we compute the total energy and the corresponding energy density. We then toss a new disorder realization and repeat the process $N_r$ times. We finally compute the average energy density $h$ and its standard deviation. The resulting dependence $h(a)$ is shown with red circles in Fig.~\ref{fig:phasediagram}a, where the error bars are the standard deviation. We find that for nonzero $W$ (here $W=4$) the curve $h(a)$
shifts to lower energies as compared to the ordered $W=0$ curve.

Averaging both sides of Eq.~(\ref{eq:a_generalrule}) over all sites in an infinite system yields the chemical potential
\begin{equation}\label{eq:muavg}
\mu= g {a} -J \bar{\zeta},
\end{equation}
with the average GS correlation field value $\bar{\zeta}=\langle \zeta_\ell \rangle$. Inserting Eq.~(\ref{eq:muavg}) into Eq.~(\ref{eq:a_generalrule}) leads to
\begin{equation}\label{eq:gglsquare}
    g G_\ell^2=ga-\epsilon_\ell +J \delta\zeta_\ell \geq 0
\end{equation}
where $\delta \zeta_\ell=\zeta_\ell-\overline{\zeta}$ describes the fluctuations of the GS correlation field.

\begin{figure}[tp]
\centering
\includegraphics[width = 8cm]{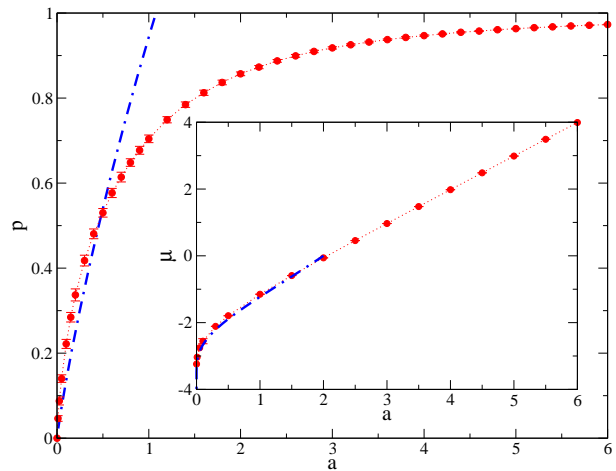}
\caption{The ground state participation number density
$p$ versus the particle density $a$ (red circles).
Error bars represent the standard deviation of the data.
Dotted lines connect the data and guide the eye.
The blue dashed-dotted line is obtained from Eq.~(\ref{eq:pa}).
Inset:
The chemical potential $\mu$ versus $a$  (red circles). Error bars represent the standard deviations of the data.
Dotted lines connect the data and guide the eye. The blue dashed-dotted line is obtained from Eq.~(\ref{eq:a_mikhail2}).
\\
$N=1000$, $N_r=100$, $W=4$, $g=1$, $J=1$ for both plots.
}
 \label{fig:p_a_mu}
\end{figure}

\begin{figure}[tp]
\centering
\includegraphics[width = 8cm]{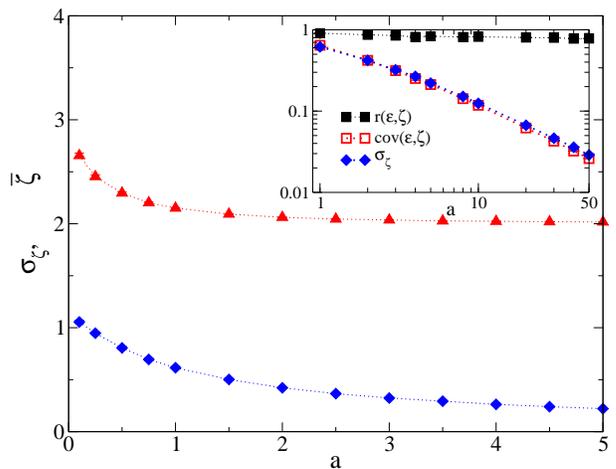}
\caption{The average GS correlation field value
$\bar{\zeta}$ (red filled triangles) and its variance $\sigma_{\zeta}$ (blue filled diamonds) vs $a$.
Inset: $\sigma_{\zeta}$ (blue filled diamonds), $\mathrm{cov}(\epsilon,\zeta)$ (red open squares) and $r(\epsilon,\zeta)$ (black filled squares) versus
$a$ on logarithmic scales.
$N=1000$, $N_r=100$, $W=4$, $g=1$, $J=1$ for both plots.
Connecting dotted lines guide the eye.
}
 \label{fig:zeta}
\end{figure}

The inhomogeneity of the ground state is measured with its participation number density
\begin{equation}
    p =\frac{N a^2}{\sum_{\ell}^N {|G_{\ell}|}^4}\;,\; 0 \leq p \leq 1\;.
    \label{pndensity}
\end{equation}
For $p \rightarrow 1$ the GS turns homogeneous, while $p \rightarrow 0$ indicates a sparse GS.
The numerical evaluation of $p(a)$ is shown in Fig.~\ref{fig:p_a_mu}. We find that $p$ takes small values in the
weak interaction regime as seen also by a GS realization in Fig.~\ref{fig:phasediagram}b. In the strong interaction
regime, $p$ tends to one, thus
indicating a more and more homogeneous GS distribution as seen also by a
GS realization in Fig.~\ref{fig:phasediagram}c. The inset in Fig.~\ref{fig:p_a_mu} shows the computed $\mu(a)$
dependence. In particular, $\mu(a\rightarrow 0) \rightarrow -2J-W/2= -4$, while $\mu=ga-2J$ in the strong interaction regime.

The GS correlation field $\zeta_l$ characteristics are captured by its average $\bar{\zeta}$ and standard deviation $\sigma_{\zeta}$. Their dependence on $a$ is
represented in Fig.~\ref{fig:zeta}.
In the strong interaction regime, it
quickly approaches its asymptotic value $\bar{\zeta}=2$. The standard deviation $\sigma_{\zeta} \sim 1/a$ tends to zero in the same strong interaction limit (see also inset in Fig.~\ref{fig:zeta}). The same inset also shows that the
covariance $\mathrm{cov}(\epsilon,\zeta)=\langle\epsilon_\ell\delta\zeta_\ell\rangle$ is tending to zero in a similar way in the
strong interaction regime. However, the relative correlation $r(\epsilon,\zeta)=\mathrm{cov}(\epsilon,\zeta)/\sigma_\epsilon\sigma_\zeta\approx 0.8$
appears to be almost constant and nonzero in the limit of strong interactions, indicating that whatever weak the
GS correlation field fluctuations may become, they still carry some nontrivial correlations with the disorder field.

\subsection{Weak interaction regime}

In the low particle density regime, i.e. as  $ga \ll W$ the ground state $G_\ell$ is a strongly inhomogeneous one. In order to satisfy Eq.~(\ref{eq:a_generalrule}), it has to be contained by a  large amount of well-separated particle clusters of size $L \geq 1$
(see Fig.~\ref{fig:phasediagram}b). The particles concentrate in the deepest minima of the on-site energy landscape. Let us perform a quantitative analysis of these so-called Lifshits states \cite{LuganPRL98}. Let $\rho_1$ be the probability of finding a particle on site $\ell$. The on-site energy $\epsilon_\ell$ has to satisfy the condition  $\epsilon_\ell<\mu+J\zeta_\ell$. We use the probability $\rho_\mathrm{c}(\mu)$ to obtain particles in a
connected cluster of size $L$ \cite{kramer}
\begin{equation}\label{eq:a_mikhail}
\rho_\mathrm{c}(\mu)= {\rho_1}^L= \left[ \int_{-W/2}^{\mu+2J} \rho d\epsilon \right]^L= {\left[ \frac{\mu+2J+W/2}{W} \right]}^L.
\end{equation}
The cluster size $L$ is obtained as follows. Inside a  cluster, the ground state $G_\ell$ satisfies the linearized Eq. (\ref{eq:eq4}) with minimal  on-site energy, $\epsilon_\ell=-W/2$. Therefore, $G_\ell \simeq \sin (q_0 \ell)$, where $q_0$ is determined as
$q_0= \sqrt{(\mu+2J+W/2)/J}$.
Taking into account that $G_\ell \simeq 0$ outside the cluster,
we obtain the explicit expression
\begin{equation}\label{eq:L_mikhail}
L(\mu)= \pi \sqrt{\frac{J}{\mu+2J+W/2}}.
\end{equation}

The total norm  can be approximated as the product of
the total number of clusters $N_0$, multiplied by their amplitudes $A_0$, and the cluster size $L$:
 \begin{equation}\label{eq:a_small}
\mathcal{A}= \sum_\ell {\psi_\ell}^2=N_0A_0L=N\rho_\mathrm{c}(\mu) A_0 L.
\end{equation}

Using Eq.~(\ref{eq:a_small}) it follows
\begin{equation}\label{eq:a_mikhail2}
g a=g A_0 L \rho_\mathrm{c}(\mu)\approx 2 {\left[ \frac{\mu+2J+W/2}{W} \right]}^L,
\end{equation}
where $g A_0 L \approx 2$, as found numerically.
Substituting (\ref{eq:L_mikhail}) in (\ref{eq:a_mikhail2}) and solving the resulting transcendent equation we obtain the dependence of the chemical potential on the norm density $\mu(a)$.  This dependence is plotted in the inset of Fig.~\ref{fig:p_a_mu} by the dashed-dotted line and shows very good agreement with the numerically computed $\mu(a)$ dependence.


Similarly, we obtain the dependence of  the ground state participation number density $p$ on the chemical potential $\mu$ as
\begin{equation}\label{eq:pa}
p= \frac{(\sum_\ell {{\psi_\ell}^2})^2}{\sum_\ell {\psi_\ell}^4}= \frac{({A_0} L   {\rho_\mathrm{c}(\mu)} N)^2}{ {A_0}^2 L  \rho_\mathrm{c}(\mu)N}=N \rho_\mathrm{c}(\mu)L.
\end{equation}
With Eq. (\ref{eq:a_mikhail2}) it follows
\begin{equation}\label{eq:analytical-p}
p=N\frac{L}{2}ga.
\end{equation}
Using (\ref{eq:L_mikhail}) and (\ref{eq:a_mikhail2}) we obtain the $p(a)$ dependence  plotted in Fig.~\ref{fig:p_a_mu} which shows good agreement with the numerically computed result.

Finally, we can compute the ground state energy density by
substituting (\ref{eq:pa}) into the exact relation $h=\mu a -{ N g a^2}/{2 p}$:
\begin{equation}\label{eq:hfistul}
h=\mu a-\frac{g a^2 }{2\rho_\mathrm{c}(\mu)L}=\left(\mu-\frac{1}{L} \right)a.
\end{equation}
This dependence is plotted in Fig.~\ref{fig:phasediagram} by the dashed-dotted blue line and shows very good agreement
with the numerical result for small values of $a < 1$.

\subsection{Strong interaction regime}
In the strong coupling regime, we have $\max \{ W,J\}\ll ga$.
First, we  use Eq.~(\ref{eq:gglsquare}) when $J/(ga)\ll 1$
\begin{align}
{G_\ell}\simeq\sqrt{a-\frac{\epsilon_l }{g}}.\label{gglsquare1}
\end{align}
Using Eqs.~(\ref{eq:H}) and (\ref{eq:psilt}) and averaging over all sites with respect to disorder yield
\begin{equation}\label{eq:h_avg1}
h=  \left\langle \epsilon_\ell G_\ell^2 \right \rangle +\frac{g}{2} \left \langle G_\ell^4 \right\rangle -2J  \langle G_\ell G_{\ell+1}\rangle. 
\end{equation}
This finally leads to the energy density of the ground state in the strong interaction regime (see also
Appendix \ref{APPAGS})
\begin{equation}\label{eq:h_large_a}
h\approx  \frac{g}{2}\left(a^2-\frac{w^2}{12}\right) -\frac{8J}{9 w^2}\left[\left(a+\frac{w}{2}\right)^{3/2}-\left(a-\frac{w}{2}\right)^{3/2}\right]^2,
\end{equation}
where we denote $w=W/g$.
The corresponding black dashed curve for $W=4$ in Fig.~\ref{fig:phasediagram} shows very good
agreement with the numerically computed ground state line $h(a)$ in the strong interaction regime.

Note that in the strong-coupling regime we can further expand Eq.~(\ref{gglsquare1}) in small parameter $W/(ga)$
\begin{align}
{G_\ell} \simeq \sqrt{a} \left[ 1 - \frac{\epsilon_l }{2ga} 
  \right].\label{gglsquare}
\end{align}
Then the resulting ground-state energy is very close to that of Eq.~(\ref{eq:h_large_a}).

\section{Elementary excitations}

We consider weak (small amplitude) excitations above the ground state by introducing a small perturbation to the ground state $G_\ell$:
\begin{equation}\label{eq:delta}
\psi_{\ell} (t) =(G_\ell+\delta_\ell(t)) e^{-i\mu t}.
\end{equation}
Linearizing the equations of motion (\ref{eq:eom}) with respect to $\delta$, we arrive at
\begin{equation}
i\dot{\delta_{\ell}}=(\epsilon_{\ell}-\mu) \delta_{\ell} + g {G_{\ell}}^2( \delta_{\ell}^* + 2 \delta_{\ell}) - J(\delta_{\ell+1} + \delta_{\ell-1}) \label{eq:idotdelta_l}.
\end{equation}
With the choice
\begin{equation}\label{eq:delta_l1}
\delta_{\ell}(t)= \chi_{\ell} e^{-i\lambda t} - {\Pi_{\ell}}^* e^{i\lambda t}
\end{equation}
and the use of Eq.~(\ref{eq:a_generalrule})  we obtain
the BdG eigenvalue equations for the two interacting excitation fields
$\chi$ and $\Pi$ as
\begin{align}\label{eq:chipi}
\lambda \chi_{\ell} =& J \zeta_{\ell} \chi_{\ell} -J(\chi_{\ell+1}+\chi_{\ell-1}) -g G_{\ell}^2(\Pi_{\ell} -\chi_{\ell}), \\
 \lambda \Pi_{\ell} =&- J \zeta_{\ell} \Pi_{\ell} +J(\Pi_{\ell+1}+\Pi_{\ell-1})+ g G_{\ell}^2( \chi_{\ell} -\Pi_{\ell}). \nonumber
\end{align}
We stress that the RHS of the above BdG equations contains only the GS field $G_l$ as input.
The BdG equations are particle-hole symmetric $\left[\lambda_\nu, \left\{{\chi_\ell}^\nu,{\Pi_\ell}^\nu\right\}\right] \longleftrightarrow \left[-\lambda_\nu,\left\{{\Pi_\ell}^\nu,{\chi_\ell}^\nu\right\}\right]$ ($\nu$ is the mode number).
This yields the solution $\chi_\ell =\Pi_\ell \propto G_l$ for $\lambda_\nu = 0$ as follows also readily from inspecting (\ref{eq:chipi}), independently
of the choice of the GS field.
The eigenvector to $\lambda=0$ can be also obtained from the observation that $\delta_\ell$ is time-independent by Eq.~(\ref{eq:delta_l1}). Then it follows from Eq.~(\ref{eq:delta}) that $\delta_\ell$  is proportional to $G_\ell$. Due to this proportionality,
the BdG eigenvector to $\lambda=0$ is delocalized in space, just as $G_\ell$ is for any finite
norm density $a$ (see also
Ref.\cite{Fetter1971}).

Next we use the decomposition
$\chi_\ell=(S_\ell+D_\ell)/2$ and $\Pi_\ell=(S_\ell-D_\ell)/2$ 
we arrive at the (still) exact set of equations
\begin{align}
\lambda S_{\ell}&=(J\zeta_\ell+2g G_{\ell}^{2}) D_{\ell} -J(D_{\ell+1}+D_{\ell-1}), \label{Sl} \\
\lambda D_{\ell}&=J\zeta_\ell S_{\ell}- J(S_{\ell+1}+S_{\ell-1}) \label{Dl}.
\end{align}
Inserting (\ref{Dl}) into (\ref{Sl}) yields
\begin{widetext}
\begin{align}
\label{onlyS}
\frac{\lambda^2}{J} S_{\ell} = (J\zeta_\ell+2g G_{\ell}^{2})(\zeta_\ell S_{\ell}- S_{\ell+1}-S_{\ell-1}) - J(\zeta_{\ell+1} S_{\ell+1} +
\zeta_{\ell-1} S_{\ell-1} - S_{\ell-2} - 2S_\ell - S_{\ell-2} ) \;.
\end{align}
\end{widetext}

BdG modes to nonzero values of $\lambda$ are expected to be Anderson localized due to
the presence of disorder and the one-dimensionality of the system.

\subsection{Numerical results}

\begin{figure}[tp]
\includegraphics[width=0.5\textwidth]{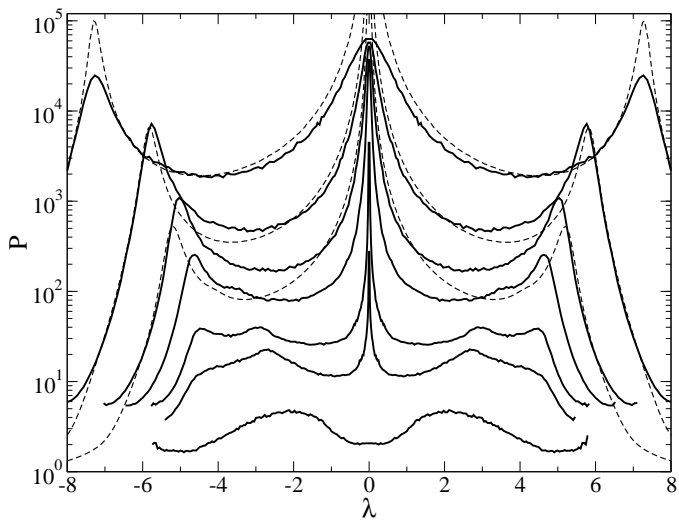}
\caption{
The bin average participation number $\bar{P}$ of BdG modes versus the energy $\lambda$.
Solid lines - numerical computation with $N=10^5$, $W=4$ and
$a=0, 0.5, 1, 2, 3, 5, 10$ from bottom to top, and $N_r=3$. Thick dashed lines - transfer matrix calculation results of the localization length $1.5 \times \xi(\lambda)$ for $a=3,5,10$ with $10^8$ iterations of (\ref{onlyS}) (see text for details). Note that all localization length data are multiplied with one and the same scaling factor $1.5$.
} \label{fig:PlaN1d5}
\end{figure}

We numerically calculate the participation number of each mode as
\begin{equation}\label{eq:P_eigen}
P_\nu= \frac{\left(\sum_{\ell}^N  n_{\ell,\nu}  \right)^2}{\sum_{\ell}^{N}n_{\ell,\nu}^2} \; , \; n_{\ell,\nu} = |\chi_{\ell,\nu}|^2 +|\Pi_{\ell,\nu}|^2
\end{equation}
where $\nu=1,\dots,2N$ is the mode number. We divide the $\lambda$-axis into small bins of size 0.05,  and average the
participation numbers in each bin to obtain the dependence $\bar{P}(\lambda)$. The system size is $N=10^5$, and we used three disorder
realizations (except for $a=10$ with only one disorder realization).
The resulting curves are plotted in Fig.~\ref{fig:PlaN1d5} for different norm densities and $W=4$. We observe symmetric
curves $\bar{P}(\lambda)=\bar{P}(-\lambda)$ due to the above-mentioned particle-hole symmetry of the BdG
eigenvalue problem. All curves show a clear divergence $\bar{P} (|\lambda| \rightarrow 0) \rightarrow \infty$ which is
only limited due to finite-size effects. This divergence agrees with the above result
that the BdG mode at zero energy $\lambda=0$ must be delocalized and thus have an infinite participation number. The divergence has been addressed in a number of publications \cite{Pavloff,Fetter1971,Chalker,Deng,Fontanesi_PRA,Fontanesi_PRL} which results in $\bar{P} \sim 1/|\lambda|^{\alpha}$ with $\alpha=2$ in the strong interaction regime \cite{Pavloff,Ishii,Ziman}.

The participation numbers at nonzero energies show
an expected finite height peak in the weak interaction regime which is a peak continuation
from the zero interaction limit (Fig.~\ref{fig:PlaN1d5}). In that limit, the BdG eigenvalue equations decompose into two copies
of the tight-binding chain with onsite disorder. The largest localization length and participation number are then obtained in the centers of their spectra $\lambda=\pm \mu$ which host the largest density of states \cite{Kramer93}. Upon crossing over from the weak interaction to the strong interaction regime, we observe a second peak developing at larger absolute $\lambda$ values, which has a finite height, but which appears to grow swiftly (Fig.~\ref{fig:PlaN1d5}). This new side peak results in an unexpected enhancement of the participation number, localization length, and size of BdG modes at finite energies $\lambda$ in the strong interaction regime.

\subsection{Finite momentum localization length singularity for strong interactions}

\begin{figure}[tp]
\includegraphics[width=0.5\textwidth]{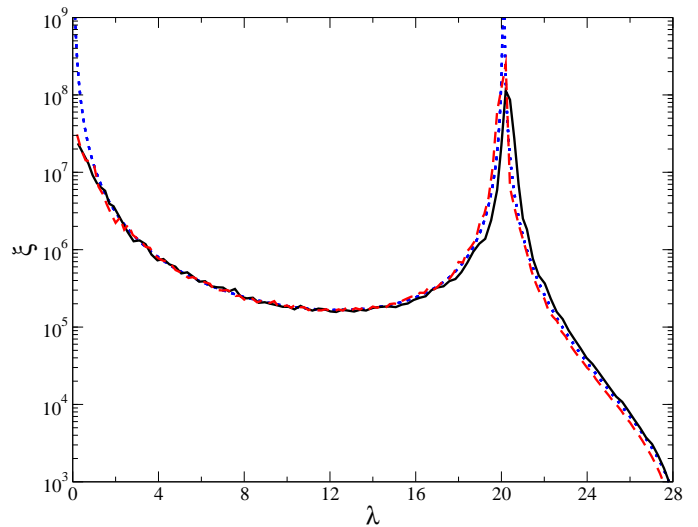}
\caption{
The localization length $\xi(\lambda)$ for $a=100$, $g=J=1$, $W=4$.
Black solid line: exact BdG equations (\ref{onlyS}), approximate GS field (\ref{gglsquare}), transfer matrix calculation results with $10^8$ iterations. Red thick dashed line: approximate BdG equations
(\ref{Slap}), approximate GS field (\ref{gglsquare}), transfer matrix calculation results with $10^8$ iterations. Blue thick dotted line - analytical result (\ref{invloc_fin}) with (\ref{Edef}).
}
\label{largea}
\end{figure}

In order to analytically assess the observed side peak of the BdG modes in the strong interaction regime
$g a \gg W$, we use the exact equations (\ref{onlyS}) with the approximated GS field (\ref{gglsquare}) and compute the localization length
$\xi(\lambda)$ with a transfer matrix method (see  Appendix \ref{LLTMM}). The resulting curves are plotted in Fig.\ref{fig:PlaN1d5} for $a=3,4,5,10$
and show almost full quantitative agreement with the numerical results from the exact equations and the numerically exact GS for $a=10$, while the
agreement is less quantitative but still qualitative as the value of $a$ is reduced. Therefore we can use the
approximate GS field (\ref{gglsquare}) with the exact equations (\ref{onlyS}) as a reliable reference for even larger values of $a > 10$, which are
not accessible by brute force numerical computations. The resulting dependence $\xi(\lambda)$ is shown in Fig.~\ref{largea} for $a=100$. The
side peak is not only remaining in place but is also increasing its height relative to the background.

In the strong interaction limit, $ga \gg W$, deeper insight on the localization properties of BdG modes can be obtained. We simplify the exact Eq. (\ref{onlyS}) keeping only the leading order term on the RHS of (\ref{onlyS}), and arrive at
\begin{align}\label{Slap}
  \frac{\lambda^2}{J}S_\ell=2g G_{\ell}^{2}(\zeta_{\ell} S_{\ell} -S_{\ell-1} -S_{\ell+1}).
\end{align}
Note that $\lambda=0$ and $S_l=G_l$ is still a valid solution.
We again compute the localization length $\xi(\lambda)$ using (\ref{Slap}) and the GS field approximation (\ref{gglsquare}). The resulting curve
for $a=100$ agrees quantitatively with the exact equation result in Fig.~\ref{largea}, confirming the validity of our equation approximation.

Defining the dimensionless energy $E$ as $E=\lambda^2/(2gaJ)$
we cast equation (\ref{Slap}) into the standard form
\begin{align}\label{stand}
 (\tilde{E}+\varkappa_\ell)S_\ell=S_{\ell-1} +S_{\ell+1}
\end{align}
with
\begin{align}
\tilde{E}=\langle \zeta_\ell\rangle-E\langle a/G_{\ell}^{2}\rangle\;,\;
\varkappa_\ell=\zeta_\ell-E\frac{a}{G_{\ell}^{2}}-\tilde{E}.\label{kapp}
\end{align}
In the strong interaction regime $ga \gg W$, $\tilde{E}=2-E$ and the perturbing random potential $\varkappa_\ell$ is small,
with its expectation value being zero: $\langle\varkappa_\ell\rangle =0$.
The disorder field $\epsilon_\ell$ is uncorrelated at different sites: $\langle\epsilon_n\epsilon_m\rangle =\delta_{nm} W^2/12$.
This holds as well for the ground state field (\ref{gglsquare}) in the strong coupling regime. However, the ground state correlation field $\zeta_\ell$ has a finite range of correlations due to the presence of nearest neighbor terms $G_{\ell\pm1}$ in its definition (\ref{eq:a_generalrule}). As a consequence, the random potential $\varkappa_\ell$ is also correlated.
Anderson localization with correlated disorder was studied in many publications (see Ref.~\cite{Lifshits88:book} for continuum models and Refs.~\cite{Griniasty88,Luck89,Izrailev99,Titov05} for lattice models).
The localization length of model (\ref{stand}) is calculated following Sec.~5.2.1 of the review \cite{Izrailev12} (see Appendix \ref{LLCD} for details):
\begin{align}\label{invloc_fin}
\xi=\frac{96g^2a^2}{W^2}\frac{4-E}{E(2-E)^2}\;.
\end{align}
In the vicinity of $E=\lambda^2/(2gaJ)=0$, we re-obtain the known localization length divergence ${\xi}\simeq \frac{192g^3a^3 J}{W^2\lambda^2}$.
Notably, we discover an additional divergence of the localization length at finite
energy $E=2$, i.e.
$\lambda = \pm 2\sqrt{gaJ}$,
as
\begin{align} \label{sidepeak}
\xi \simeq \frac{24g^3a^3J}{W^2 (2\sqrt{gaJ} \pm \lambda)^2}\;, \; |\lambda \pm 2\sqrt{gaJ}| \ll 2\sqrt{gaJ} \;.
\end{align}
The above singularity is the explanation for the observed side peak. We plot (\ref{invloc_fin}) in
Fig.~\ref{largea} for $a=100$ and find quantitative agreement with the localization length data from transfer matrix evaluations of the exact and approximate equations while using the approximate GS field dependence as induced by the disorder.

\subsection{Generalizations}

Let us generalize to any lattice dimension with some hopping network or generalized discrete Laplacian:
\begin{equation}\label{eq:eomGen}
i\dot{\psi_{\ell}}=\epsilon_{\ell} \psi_{\ell} +  g|\psi_{\ell}|^2\psi_{\ell}-\mathcal{D}(\psi_\ell)
\end{equation}
The discrete Laplacian
\begin{equation}\label{DL}
\mathcal{D}(\psi_\ell) = \sum_m J(\ell-m)\psi_m\;.
\end{equation}
We assume $J(m) \geq 0$ to ensure the nonnegativity of the ground state field $G_l$. Note that the Hamiltonian (\ref{eq:H}) is obtained with the choice  $J(m)=J(\delta_{m,1}+\delta_{m,-1})$.
It follows from the definition (\ref{DL}) that
\begin{equation}\label{eq:gglsquare_gen}
    g G_\ell^2=ga-\epsilon_\ell + \delta\hat{\zeta}_\ell \geq 0
\end{equation}
and
\begin{equation}\label{zetal}
\hat{\zeta}_l = \frac{1}{G_l} \mathcal{D}(G_l)
\end{equation}
Note that with this definition the field $\hat{\zeta}_l$ includes the strength of the hopping network, as opposed to previous notations.

The \emph{exact} equations for $S_\ell$ and $D_\ell$ take the form
\begin{align}
\lambda S_{\ell}&=(\hat{\zeta}_\ell+2g G_{\ell}^{2}) D_{\ell} -\mathcal{D}(D_l)  \label{Sl2}\\
\lambda D_{\ell}&=\hat{\zeta}_\ell S_{\ell}- \mathcal{D}(S_l)  \label{Dl2}.
\end{align}
The approximate expression for the field $G_\ell$ in the strong-coupling case is still given by Eq. (\ref{gglsquare}), with all corrections due to the change in the hopping network and even the dimensionality appearing in higher-order corrections.

Since both $\hat{\zeta} \sim J$ and $\mathcal{D} \sim J$, we arrive at the generalized strong interaction BdG equations similar to the
above considered one-dimensional case with nearest neighbor hopping as
\begin{align}\label{Slap2}
  \lambda^2 S_\ell=2g G_{\ell}^{2} \left( \hat{\zeta}_{\ell} S_{\ell} -\mathcal{D}(S_l)\right)  .
\end{align}
Equations (\ref{gglsquare}), (\ref{DL}), (\ref{zetal}), and (\ref{Slap2}) constitute the generalization of the BdG equations in the strong coupling limit to any lattice dimension and hopping
network. It remains to be studied whether these equations also result in a strong enhancement of transport properties of disordered BdG modes
at certain finite energies and momenta due to ground state correlations.

\section{Conclusion}

We have studied in detail the properties of the ground state and small amplitude BdG excitations of the one-dimensional Gross--Pitaevskii lattice model in the presence of spatial disorder. On the numerical side, we computed the ground state energy density $h$ as a function of the norm (particle) density $a$
throughout the weak and strong interaction regimes. We applied perturbation approaches to both regimes and obtained analytical approximations of $h(a)$,
the chemical potential, and the participation number density of the ground state which characterizes its spatial distribution properties. The obtained analytical results agree quantitatively with full numerical computations.

We then proceeded to numerically compute the localization properties of small amplitude excitations above the ground state, which are coined Bogoliubov-de Gennes modes. We observe a divergence of their localization length for zero energy in full accord with previous publications. However, we also find
an anomalous enhancement of the localization length of excitations in a side peak for finite energies in the strong interaction regime. We perform a systematic perturbation approach which results in approximate eigenvalue equations which are valid in the strong interaction regime. That eigenvalue problem corresponds to
a one-dimensional tight-binding chain with nearest-neighbor hopping and correlated on-site disorder. We derive analytical expressions for the localization length as a function of energy. We then finally obtain a singularity and length divergence at a finite energy, which precisely corresponds to the numerically
observed side peak. Therefore we conclude that a weakly excited disordered condensate in the regime of strong interaction will allow for almost ballistic transport of excitations for selective finite energies and momenta. We also generalize the strong interaction equations for Bogoliubov-de Gennes modes for
more complicated and higher dimensional networks.

{\bf Acknowledgements}
YK thanks Tilen Cadez for help with the transfer matrix coding, and  A. Andreanov, C. Danieli, I. Vakulchyk, and S. Gundogdu for insightful discussions. This work was  supported  by  IBS-R024-D1. M.V.F. acknowledges partial financial support of the Ministry of Science and Higher Education of the Russian Federation in the framework of the State Program (Project No. 0718-2020-0025). A.~Yu.~Ch. acknowledges support from the JINR–IFIN-HH
projects and thanks the hospitality of the IBS Center for Theoretical Physics of Complex Systems.

\appendix

\section{Numerical Details on the Ground State Calculation}
\label{sec:numGS}
The wavefunction amplitude is defined as $\psi_\ell=\sqrt{a_\ell}e^{i\phi_\ell}$, where  $a_\ell$ is the local norm density, and $\phi_\ell$ is the phase of the complex order parameter at each site with $0\leq \phi_\ell\leq 2\pi$. It follows from Eq. (\ref{eq:H}) that the ground state is characterized by a vanishing phase difference between neighboring sites, $\phi_\ell=\phi$, and all the phases evolve in time keeping their phase difference zero [Eq. (\ref{eq:psilt})]. Without loss of generality, we set $t=0$ when all phases are equal to zero and define $G_\ell$ as a real variable.

Fixing the disorder realization and the desired average norm density $a$, we start iterating with a real random initial guess for $G_\ell$ in order to minimize the real function of $H$.
The iteration is performed on five neighboring sites at a time using a function minimization algorithm \cite{lagarias}, along with the normalization of $G_\ell$ according to $\sum_\ell G_\ell^2=N a$. We shift the window of minimization by one site until all sites of the system are iterated once. We repeat this procedure 10-40 times until full convergence to the minimum energy is reached with a tolerance $<10^{-15}$. We then test the numerical ground state $G_\ell$ such that the standard deviation of $\mu$ for each  site found by  Eq.~(\ref{eq:a_generalrule})  is  much  smaller  than  its  mean  for  each realization of a system size $N$.

\section{Ground State Averaging for Strong Interactions}
\label{APPAGS}

For $a \geq W/2g$, we can define the average interaction of the neighboring sites as

\begin{equation}
 \langle G_\ell G_{\ell+1} \rangle= \left\langle \sqrt{\left(a-\frac{\epsilon_{\ell}}{g}\right)\left(a-\frac{\epsilon_{\ell+1}}{g}\right)} \right\rangle.
\end{equation}

Different sites are uncorrelated, i.e. for arbitrary $f(x)$:
$    \langle f(\epsilon_i)f(\epsilon_j)\rangle=\langle f(\epsilon_i)\rangle\langle f(\epsilon_j)\rangle$,
for $i\neq j$. Then
\begin{equation}\label{eq:fsquare}
    \frac{1}{N}\sum_i \langle f(\epsilon_i)f(\epsilon_j)\rangle = \left(\int d\epsilon \rho(\epsilon)f(\epsilon)\right)^2
\end{equation}

By employing $f(\epsilon)=\sqrt{a-{\epsilon}/{g}}$ in Eq.~(\ref{eq:fsquare}), we get
\begin{align}
&\left\langle \sqrt{\left(a-\frac{\epsilon_{\ell}}{g}\right)\left(a-\frac{\epsilon_{\ell+1}}{g}\right)} \right\rangle = \left[\frac{1}{W}\int_{-\frac{W}{2}}^{\frac{W}{2}}d\epsilon \sqrt{a-\frac{\epsilon}{g}}  \right]^2 \nonumber \\
 &= \left\{ \frac{2}{3w}\left[\left(a+\frac{w}{2}\right)^{3/2}-\left(a-\frac{w}{2}\right)^{3/2}\right] \right\}^2,
  \label{eq:T}
\end{align}
where $w=W/g$.

\section{Localization Length : Transfer Matrix Method}
\label{LLTMM}

We solve Eq.~(\ref{onlyS}) with the approximate GS field (\ref{gglsquare}) with the transfer matrix method:


\begin{equation}\label{4by4}
\begin{bmatrix}
S_{\ell+2}  \\
S_{\ell+1} \\
S_\ell \\
S_{\ell-1}
\end{bmatrix} =
T_\ell
\begin{bmatrix}
S_{\ell+1}  \\
S_{\ell}\\
S_{\ell-1} \\
S_{\ell-2},
\end{bmatrix}
\end{equation}

where the transfer matrix
\begin{equation}
T_\ell=
\begin{bmatrix}
u_\ell+\zeta_{\ell+1} & {\lambda^2}/{J^2}-2-u\zeta_\ell & u_\ell+\zeta_{\ell-1} & -1 \\
 1 & 0 & 0 & 0 \\
0  & 1 & 0 & 0 \\
0 & 0 & 1 & 0
\end{bmatrix},
\end{equation}
where $u_\ell=2g G_\ell^2/J +\zeta_\ell$.
Following \cite{Slevin_2014}, we start the transfer matrix multiplication with
\begin{equation}
Q_0=
\begin{bmatrix}
1 & 0  \\
0 & 1 \\
0 & 0 \\
0 & 0
\end{bmatrix}
\end{equation}
matrix  with orthogonal columns, and multiply it by the transfer matrices $T_1 T_2 \dots T_N$.
To control the round-off error, after each $q=5$ iterations (number of multiplications by the transfer matrix $\hat{T}$), we apply QR decomposition which gives two vectors: a normalized $2 \times 4$  matrix $Q_j$, and a $2\times2$ upper triangular matrix $R_j$:
\begin{equation}
Q_{j} R_{j} =T_{({j-1})q+1}  \dots T_{jq}  Q_{j-1}  \quad (j=1, \dots, N/q)
\end{equation}
The smallest positive Lyapunov exponent can be estimated by
\begin{equation}
\tilde{\gamma}_2= \frac{1}{N} \sum_{j=1}^{N/q} \ln{R_j(2,2)}
\end{equation}
in the limit $N\rightarrow \infty$. Here, in each QR factorization step, we store the second diagonal element of $R_j$ which has the smaller positive Lyapunov exponent.
Practically, we need to use large but finite number of iterations: $N=10^8$.
\begin{equation}
\xi= 1/\gamma_2.
\end{equation}

In addition, Eq.~(\ref{Slap}) can be solved by transfer matrix method:

\begin{equation}\label{4by4}
\begin{bmatrix}
S_{\ell+1} \\
S_\ell
\end{bmatrix} =
T_\ell
\begin{bmatrix}
S_{\ell}\\
S_{\ell-1}
\end{bmatrix},
\end{equation}

where
\begin{equation}
T_\ell=
\begin{bmatrix}
\zeta_\ell -\lambda^2/2 J g G_\ell^2 & -1 \\
 1 & 0
\end{bmatrix}.
\end{equation}

We start the transfer matrix multiplication with
\begin{equation}
V_0=
\begin{bmatrix}
1   \\
0
\end{bmatrix}
\end{equation}
vector, and multiply it by the transfer matrices: $V_j = T_{j}  V_{j-1}  \quad (j=1, \dots, N)$. 
To deal with the round-off error, after every $5$ iterations, we normalize the vector $V_j$ and estimate the smallest positive Lyapunov exponent by
\begin{equation}
\tilde{\gamma}_1= \frac{1}{N} \sum_{j=1}^{N/q} \ln{\Vert{V_j}\Vert}, \qquad
\end{equation}

After $N=10^8$ iterations, we find the localization length $\xi$ as
\begin{equation}
\xi= 1/\tilde{\gamma_1}.
\end{equation}

\section{Localization length calculation details in the strong interaction regime}
\label{LLCD}

In this appendix, we outline the computation of the localization length in the strong interaction regime as given by Eq.~(\ref{invloc_fin}). We start from Eqs.~(\ref{stand},\ref{kapp}).
In leading order $1/(ga)$ we obtain $\tilde{E}=2-E$ and the onsite disorder potential
\begin{align}
\varkappa_l = \frac{\epsilon_l(2-2E)-\epsilon_{l-1} -\epsilon_{l+1}}{2ga}\;.
\end{align}
We define the onsite disorder correlation function
\begin{align}
K(n-m) =K(m-n) =\langle\varkappa_n\varkappa_m\rangle \;.
\end{align}
The range of the correlations is finite because $K(\ell)$ takes non-zero values only for $\ell=0,\pm 1,\pm 2$:
\begin{eqnarray}
K(0)&= & \langle \varkappa_l^2 \rangle = \frac{W^2}{48g^2a^2}\left( (2-2E)^2 +2 \right)\;,
\\
K(1)&= & \langle \varkappa_l \varkappa_{l+1} \rangle = \frac{W^2}{24g^2a^2}(2E-2)\;,
\\
K(2) &= & \langle \varkappa_l \varkappa_{l+2} \rangle = \frac{W^2}{48 g^2 a^2}\;.
\end{eqnarray}
The Fourier transformed correlation function is then readily obtained:
\begin{widetext}
\begin{eqnarray}
\label{kq}
K(q)=\sum_{\ell} K(\ell)e^{i q \ell} =K(0)+2\sum_{\ell=1}^{\infty} K(\ell)\cos q \ell = \frac{W^2}{12 g^2a^2}\left( (1-E)^2 + \frac{1}{2} -2(1-E)\cos q + \frac{1}{2}\cos 2q \right) \;.
\end{eqnarray}
\end{widetext}
Next we compute the above expression at double argument value $K(2q)$ and then use the dispersion relationship of the homogeneous Eq. (\ref{stand})
\begin{equation}
\label{Edef}
E=2(1-\cos(q))
\end{equation}
to replace $q$ by $E$. After some additional simple algebra, the result reads
\begin{align}
\label{k2q}
K(2q) = \frac{W^2}{48 g^2 a^2} E^2 (2-E)^2 \;.
\end{align}
Anderson localization with correlated disorder was studied in many publications (see Ref.~\cite{Lifshits88:book} for continuum models and Refs.~\cite{Griniasty88,Luck89,Izrailev99,Titov05} for lattice models). The inverse correlation length in the model (\ref{stand}) is given by (see Sec.~5.2.1 of the review \cite{Izrailev12})
\begin{align}\label{invloc_gen_a}
  \frac{1}{\xi}=\frac{K(2q)}{8\sin^{2}{q}}.
\end{align}
 Substituting $K(2q)$ from (\ref{k2q}) into Eq.~(\ref{invloc_gen_a}) and using Eq.~(\ref{Edef}) yield
\begin{align}\label{invloc_fin_a}
\xi=\frac{96g^2a^2}{W^2}\frac{4-E}{E(2-E)^2}
\end{align}

\bibliography{1DDNLSv5}

\end{document}